\documentstyle[12pt,a4]{article}
\newcommand{\be}{\begin{equation}}
\newcommand{\ee}{\end{equation}}
\newcommand{\bdis}{\begin{displaymath}}
\newcommand{\edis}{\end{displaymath}}

\title{Isotropy, homogeneity and dipole saturation}
\author{Francesco Sylos Labini}
\date{}
\begin{document}
\maketitle
\centerline{Dipartimento di Fisica, Universit\'a di Bologna, V.le Irnerio 48}
\centerline{I-40126 Bologna, Italy.}
\centerline{and}
\centerline{\footnote{Address to be used for correspondence}
Dipartimento di Fisica, Universit\'a di Roma ``La Sapienza'', P.le
Moro 2,}
\centerline{I-00185 Rome, Italy.}
\medskip
\begin{abstract}
A distribution of points that satisfies the property of local
isotropy is not necessarily homogeneous: homogeneity is
implied by the condition of local isotropy together with
the {\em assumption of analyticity or re\-gu\-la\-ri\-ty}.
Here we show that the evidence of
dipole saturation in galaxies (and clusters) catalogues,
together with a monotone growth of the monopole,  is an
evidence of isotropy but not of homogeneity.
This is fully compatible with
a fractal structure
which has the property of
local isotropy, but it
is non-analytic and non-homogeneous.
\\
\\
{\bf Subject headings}: cosmology:theory -
galaxy:clustering - large-scale structure of the universe.
\end{abstract}

\newpage

\section{Introduction}
The property of local isotropy, required by
the Cosmological Principle
in order to avoid propositions that imply privileged
points in the Universe,
together with the {\em assumption of analyticity},
implies homogeneity (Weinberg 1972).
The Cosmological
Principle is claimed to imply homogeneity directly,
while homogeneity is not
satisfied for a
non-analytic distribution of matter, that can be locally
isotropic.
The fact that local isotropy does not imply homogeneity is well known
(Weinberg 1972): here we discuss the property of local isotropy
in the case of non-analytical distribution of matter.

Various studies (for a review see
Coleman \& Pietronero (1992)) have shown that the assumption
of analyticity (regularity at large scales)
is not supported by the experimental evidence:
in fact the analysis of CfA (for galaxies) and Abell (for clusters)
catalogues show highly irregular distributions
of matter up to
the sample limit, with fractal (multifractal
if one includes mass) properties and
without any clear tendency towards  homogenisation.

Mandelbrot (1982) has formulated
a weaker Cosmological Principle,
the so-called
Conditional Cosmological Principle, which
does not require the complete homogeneity of space
(see also Coleman \& Pietronero 1992) and it is compatible with
a fractal distribution of matter: this breaks the symmetry between
occupied and empty points but all galaxies are {\em statistically
equivalent} with respect to their
environment (local isotropy).
For a complete review of this thesis
we remand to Coleman \& Pietronero (1992), as well as for the
objections to the criticism to this analysis (see also Ribeiro 1993).
We consider here another claimed evidence for homogeneity:
the dipole saturation with depth
in ga\-la\-xies and clusters ca\-ta\-lo\-gues,
together with a monotone growth of the monopole.

We are going to show
that this is an evidence of the isotropy
of the distribution but not of homogeneity.
Since a fractal structure
satisfies the property of local isotropy,
it is fully compatible with
the evidence of dipole saturation with depth,
as well as an homogeneous distribution.

A number of authors have discussed the convergence of cosmological
dipoles using different tracers: for example
flux-weighted IRAS dipoles
(Lahav et al.1988), IRAS redshift space dipole
(Rowan-Robinson et al 1990), X-ray and optical clusters
(Scaramella et al.1990; Plionis \& Valdarnini 1991; Plionis et al 1993;
Lahav et al 1989) and optical galaxies (Hudson 1993; Lahav et al.1988).

Plionis \& Valdarnini (1991) using the combined the Abell-ACO
clusters catalogues with $\:m_{0}\leq 14$,
found that the dipole converges at
$\:d_{conv}\approx 150 h^{-1}Mpc $, while the monopole
saturates for the lacking of data for deeper depth in the catalogues,
at $\:260 h^{-1}Mpc$.
Scaramella et al (1991) using a volume
limited sample of Abell-ACO
catalogues, found that $\:d_{conv}\approx 180 h^{-1}Mpc$.
The optical and infrared galaxies catalogues also indicate that
$\:d_{conv}\approx 100 h^{-1}Mpc $ (Rowan-Robinson et al.1990;
Hudson 1993).
These analysis are problematic for the lacking of data in
several sky regions in the available samples, and for the accurate
knowledge of the luminosity function required for the data weighting
(Plionis et al 1993). In various cases however the saturation of the
dipole has been interpreted as evidence for homogeneity.

We report a number of analyses on artificial
distributions with  a priori assigned properties,
for which we have studied
the behaviour of the monopole and the dipole with sample depth.
Our aim is not to reproduce the actual distribution of galaxies by
computer models, but we intend to show with these examples that
the evidence of dipole saturation does not imply homogeneity as
it is fully compatible with a non-analytic (fractal)
distribution. Such a saturation identifies
the scale beyond
which the distribution can be considered statistically isotropic.
Even in the case of a fractal distribution at small scales
with a cut-off towards homogeneity at larger scales,
it is interesting to study
the behaviour of the dipole and the monopole with sample depth
in order to clarify the physical interpretation of an eventual dipole
saturation scale.

In Sect.2 we discuss the case of an homogeneous
sample and in Sect.3 we show the properties of
fractal and multifractal samples.
Finally we present  the conclusion
of this work with special emphasis
to the concept of local isotropy
and its implication for a fractal
distribution of galaxies and clusters.

\section{Homogeneous sample}
We begin by considering the properties of a random distribution
that is really regular and homogeneous at large scale.
An homogeneous sample was ge\-ne\-ra\-ted  with a random
number generator distributed in a large spatial volume: also
the masses are assigned by a random number generator
with an uniform distribution ranging in [0,1].
Chandrasekhar (1942) provide an analytic solution for
the distribution of the
gravitational force in a random sample of points with random masses,
in the three-dimensional euclidean space.
The analytic solution is valid only
in the ther\-mo\-dy\-na\-mi\-cal limit for the
number of points $\:N$ and the volume of the sample $\:V$
that goes to
infinity, being constant the average number density $\:<n>$.
In order to analyse the effects of a finite value of $\:N$ and a finite
volume $\:V$,
as one has in the real catalogues, we have performed some
numerical simulations to study the finite size effects
and the convergency towards the asymptotic behaviour.

In our numerical simulation (Fig.1) we have
obtained a result
in agreement with the asymptotic one for $\:N=500$ (in
cube of unitary side).
For such a sample
we evaluate the monopole $\:M(r)$ defined as:
\be
M(r)=\sum_{i} \frac{m_{i}}{|\vec{r}_{i}|^2}
\ee
and the dipole $\:\vec{D}(\vec{r})$:
\be
\vec{D}(\vec{r})=\sum_{i} \frac{m_{i}\vec{r}_{i}}{|\vec{r}_{i}|^3}
\ee

Fig.2 shows that the monopole grows linearly with sample depth.
The
scale of dipole saturation $\:d_{conv}$ (Fig.3) is quickly reached:
it represents
the scale beyond which the statistical isotropy
of distribution is
reached and the further contributions to (2)
sum to zero for symmetry reason. For an homogeneous random
sample
the density is a constant function in space, a part
from statistical fluctuations,
and it is possible to compute
analytically the monopole and the dipole from eqs.(1) and (2).
We find that the monopole grows linearly with the sample depth
as shown in the simulations beyond the  scale of order of the mean
Poisson separation. The dipole modulus converges as the
mean density became  rotationally symmetric with respect to
every point: $\:n(\vec{r})\approx n(|\vec{r}|)$.

\section{Fractal and multifractal samples}
In a fractal distribution the counting of the number of points
present within a certain volume, from every occupied point, is,
for a spherical volume:
\be
N(R)=BR^D
\ee
where $\:B$ is a constant related to the lower cut-off of
the fractal, and $\:D$ is the fractal dimension
(for a more detailed definition of fractal dimension see
Paladin \& Vulpiani 1987). If $\:D = 3$, in the three-dimensional
euclidean space, the distribution is homogeneous.
The average
density for a spherical sample of radius $\:R_{s}$, which contains
a portion of a fractal structure, is:
\be
<n>=\frac{N(R_{s})}{V(R_{s})}= \frac{3B R_{s}^{D-3}}{4\pi}
\ee
{}From (4) it follows that in a fractal structure $\:<n>$ is
not a well defined quantity, i.e. it depends from the sample
depth ($\:<n>$ is constant if $\:D=3$).
The conditional density form an occupied point is defined as:
\be
\Gamma(r)=S(r)^{-1}\frac{dN(r)}{dr} = \frac{D}{4\pi} B r^{D-3}
\ee
where $\:S(r)$ is the area of a spherical shell of radius $\:r$.
Eqs. (3) and (5) hold from
every point of the system,
when considered as origin, and this means
that every point of the system has the same type of environment:
in other words a fractal structure satisfies the property of
local isotropy.
The average number density (4) is highly singular
in every occupied point and this means
that a fractal is non-analytic (Pietronero 1986).

The distribution of matter in a sample is described by the density function:
\be
\rho(\vec{r})= \sum_{i} m_{i} \delta(\vec{r}-\vec{r_{i}})
\ee
For a fractal structure the analytic calculation of the
dipole modulus and the monopole, contrary to the trivial case
of the homogeneous sample,
is not possible anymore, because one has to know the complete
distribution of mass in space to resolve eqs. (1) and (2)
explicit.
So we have preceded
with the analysis of artificial distributions.
Fig.4  shows a simple deterministic fractal:
as the structure repeats
at different scales in a self-similar way
it is obvious that
the dipole saturates with depth  quickly
from every occupied point,
while the monopole is an increasing function of depth.
This is a particular example that shows
that isotropy and homogeneity are different properties for
non-analytic distribution.
In order to consider a more general case
we can construct a random fractal or multifractal distribution
with a random $\:\beta$ model (Paladin \& Vulpiani 1987).

We first keep the mass constant in each point and
evaluate numerically the monopole and the dipole: in
doing this we are only considering the fractal
distribution of the support without any correlation between
space and mass distribution.
In order to consider also this
correlation, found in the analysis of the CfA catalogue
(Coleman \& Pietronero 1992), we have also considered
a random multifractal
distribution. In this case we assign a mass to each point
proportional to measure associated with the fragmentation process
(Benzi et al. 1984). However there is no
sensible change in
the results for these models.
By locating the observer in a random point of the distribution
(far from the boundaries) we analyze the dipole and the monopole
dependence on sample depth. In Fig.5 and Fig.6, we show the
results for fractal respectively of dimension D=1.66 and D=1.85,
while in Fig.7  D=1.4, equal to the fractal dimension found in the
catalogues analysis (Coleman \& Pietronero 1992).

We find that the monopole is an increasing function
of sample depth: this is obviously due to the fact that the monopole
measures the total gravitational field
(as the dipole measures the gravitational force)
and it increases with the number of points:
going to larger depth, also in a fractal structure,
we are simply including more points. The dependence of
$\:M(\vec{r})$ is a power law:
\be
M(\vec{r})=B r^{\alpha}
\ee
with the exponent $\:\alpha$ that depends strongly from
the topological properties of the realization:
this is due to the fact that a fractal is characterized
by having structures at all scales and also the fluctuations
in the spatial distribution of these structures are
present at all scales. For the same reason
the saturation depth $\:d_{conv}$ of the dipole modulus
depends on the particular realization of the fractal.
In any case we can conclude that a fractal (and a multifractal)
can be
locally isotropic with a finite value of $\:d_{conv}$.
The particular values of $\:d_{conv}$ and $\:\alpha$ are
related
to morphological properties of the system that cannot
be characterized simply by the value of the fractal
dimension.

\section{Conclusion}
In the previous discussion we have stressed the fact that
the property of local isotropy, required by the Cosmological
Principle, without the assumption of analyticity, does
not imply homogeneity. A fractal is locally isotropic,
but non-analytic in every occupied point, and non-homogeneous.

We have analyzed several artificial distributions with a priori
assigned properties. For the case of an homogeneous sample
we have shown that the monopole grows linearly with
sample depth, while the dipole saturates at the scale beyond which
the isotropy is statistically reached.
In the case of a fractal and multifractal distributions,
generated with a random $\:\beta$ model, the monopole
follows a power low with exponent $\:\alpha$, and the scale
of dipole saturation  $\:d_{conv}$ is finite.
Both $\:\alpha$ and $\:d_{conv}$
depend strongly on the realization
of the fractal because it is characterized by having structures at all
scale as well as fluctuations in
the spatial distribution
of these structures: it is not possible
to relate  $\:\alpha$ and $\:d_{conv}$ simply to the value
of the fractal dimension as they depend
from the morphological properties
of the system.

In the case of cosmological dipoles observed in galaxies (and clusters)
ca\-ta\-lo\-gues, from the analysis of the convergence depth  $\:d_{conv}$,
we obtain information on
the scale beyond which the distribution of
galaxies (clusters) have reached, statistically, the isotropy.
One cannot conclude that the distribution is homogeneous
beyond  $\:d_{conv}$ because the property of local isotropy
does not imply homogeneity without assumption of analyticity:
the fact that  $\:d_{conv}$ is finite is fully compatible
with a fractal structure.

In summary we have shown that a distribution of points that
satisfies the property of local isotropy is not necessarily
homogenous: homogeneity is implied by the condition of local
isotropy together with assumption of analyticity or regularity.
A fractal structure
is locally isotropic and thus compatible with the so-called
Conditional Cosmological Principle (Mandelbrot 1982), according to which
all the points of the system are statistically equivalent.

The only way to study the homogeneity
of the large scale structures
without a priori assumption (Coleman and Pietronero 1992) is to identify
the eventual scale $\:\lambda_{0}$, if present,
beyond which the two point correlation
function

$\:G(r)=<n(r_{0}) n(r+r_{0})>$
shows an evident plateau, within the statistical fluctuations.
This means that for $\:r\geq\lambda_{0}$
$\:<n(r_{0}) n(r+r_{0})> \approx <n>^{2}$
within the fluctuations of order $\:<n>^{\frac{1}{2}}$, and a definite
average density has been reached. Clearly, as the homogeneity
implies isotropy  $\:d_{conv}\leq\lambda_{0}$, and from the
knowledge of  $\:d_{conv}$ one cannot infer anything about
$\:\lambda_{0}$.

Several authors (Juszkiewicz et al. 1990, Lahav et al. 1990, and Peacock 1992)
have considered
the growth of the dipole in the case of different power spectra
of fluctuation, under the assumption of small departures from homogeneity.
The 'toy-model' described here calls attention in
interpreting dipoles if the mean density cannot be defined
because in the case of a fractal distribution of matter, as one
has density fluctuations extended at all scale, the predictions
of the linear theory are not correct.
In particular we want to stress that if an average density cannot be defined,
in other words if there is not an experimental support for analyticity,
under the assumption that light traces mass,
no value of the cosmological density parameter
$\:\Omega_{0}$ (Peebles 1980) can be
inferred from the measurement of the dipole amplitude.

\section{Acknowledgements}
The main part of the work described here was carried out in
collaboration  with L.Pietronero, and I am very grateful to him.
I want to thank  S.Borgani, R.Scaramella,
R. Valdarnini and A.Vespignani
for the useful discussions.

\newpage

\section*{References}

\begin{itemize}

\item Benzi, R., Paladin, G.,Parisi, G. and Vulpiani, A.

1984, J.Phys. A 17,3521

\item Chandrasekhar, S. 1943 Rev.Mod.Phys. 15,1

\item Coleman, P.H. \& Pietronero, L.,1992 Phys.Rep. 231,311

\item Coleman, P.H. Pietronero, L., and Sanders, R.H.  1988, A.\&A.

245,1

\item de Vaucouleurs, G. 1970, Science 167,1203

\item Hudson, M.J. 1993 MNRAS 265,72

\item Juszkiewicz, R., Vittorio, N., Wyse, R.F.G., 1990 ApJ 349, 408

\item Lahav, O., Edge, A.C.,Fabian, A.C.,Putney, A. 1989 MNRAS 238,881

\item Lahav, O., Kaiser, N., and Hoffman, Y., 1990 ApJ 352, 448

\item Lahav, O., Rowan-Robinson M.,Lynden-Bell D., 1988 MNRAS 234,677

\item Mandelbrot, B., 1982 The fractal geometry of nature, Freeman New York

\item Paladin, G. \& Vulpiani, A.1987, Phys.Rep. 156,147

\item Peacock, J.A. 1992 MNRAS 258, 581

\item Peebles, P.J.E., 1980 The Large Scale Structures of the Universe

(Princeton Univ.Press)

\item Pietronero, L., \& Tosatti, E. (eds) 1986, Fractals in Physics

(North-Holland; Amsterdam)

\item Plionis, M., Coles, P., Catelan, P. 1993 MNRAS 262,465

\item Plionis, P. \& Valdarnini, R. 1991 MNRAS 249,46

\item Ribeiro, M.B., 1993 ApJ 415,469

\item Rowan Robinson M. et al.1990 MNRAS 247,1

\item Scaramella, R.,Vettolani, G., Zamorani, G. 1991 ApJ 376,L1

\item Weinberg, S., 1972 Gravitation and Cosmology (Weley; New York)

\end{itemize}

\newpage

\section{Figure captions}

\begin{itemize}

\item Fig.1  The solid line represents

the Chandrasekhar asymptotic solution ($\:N\rightarrow\infty$)

for the distribution  $\:W(|F|)$ of the gravitational force modulus,

 $\:|F|$, versus $\:|F|$

in a random sample with random masses in the three-dimensional euclidean

space.

The errors bars refer to

 $\:W(|F|)$ versus $\:|F|$

determined by

 an average of 10 simulations of an homogeneous sample with $\:<n>=$500.

\item Fig.2 Monopole growth with sample depth for the random samples

of Fig.1. The dependence on depth is linear.

\item Fig.3 Dipole modulus growth with sample depth for the random

sample of Fig.1. The scale of saturation is the distance beyond

which the distribution isotropy is statistically reached.

\item Fig.4 Example of simple deterministic fractal:

the same structures

repeat in a self-similar way at different scales, but the

distribution is non-analytic and non-homogeneous.

 The dipole saturates

quickly from every occupied point, and the monopole grows with

the number of points.

\item Fig.5 Monopole and dipole modulus behaviour with sample depth

for a fractal with D=1.66. One can see that the saturation depth $\:d_{conv}$

of the dipole modulus

is finite, and that the monopole

depends on depth as a power law with exponent $\:\alpha$.

\item Fig.6 Monopole and dipole  modulus behaviour with sample depth

 for a fractal with D=1.85.$\:d_{conv}$ and $\:\alpha$

are related to morphological properties of the system and depend

from the particular realization of the fractal.

\item Fig.7 Monopole and dipole  modulus behaviour with sample depth

for a fractal with D=1.4, the fractal

dimension observed in galaxies and clusters catalogues: the evidence of

dipole saturation, together with a monotone growth of the monopole,

 is not a proof of homogeneity,

as it is fully compatible

with a non-analytic (fractal) distribution.

\end{itemize}

\end{document}